\documentclass[
prl,reprint,amsmath,amssymb
]{revtex4-1}

\usepackage{graphicx} 
\usepackage{amsmath} 
\usepackage{physics}
\usepackage{dsfont}

\usepackage{hyperref} 
\usepackage[noabbrev]{cleveref} 

\usepackage[load=prefixed,load=abbr,separate-uncertainty=true]{siunitx} 

\newcommand{\Ham}{\hat{\mathcal{H}}}
\newcommand{\Fin}{\mathcal{F}}
\newcommand{\Rb}{$^{87}$Rb}
\newcommand{\sub}[1]{_{\mathrm{#1}}}
\newcommand{\subsc}[1]{_{\mathrm{\textsc{#1}}}}
\newcommand{\DTwo}{$\mathrm{D}_2$}
\newcommand{\e}{\mathrm{e}}
\newcommand{\ii}{\mathrm{i}}
\newcommand{\ie}{i.e.\@}
\newcommand{\ketrm}[1]{\ket{\mathrm{#1}}}
\newcommand{\plus}{\texttt{+}}
\newcommand{\minus}{\texttt{-}}
\newcommand{\dash}{^{\prime}}
\newcommand{\pkgfont}[1]{\texttt{#1}}

\newcommand{\diag}[1]{\mathop{\mathrm{diag}}(#1)}

\newcommand{\etal}[2]{
	\ifthenelse{\equal{#1}{}}
		{\emph{et~al.}}
		{\emph{et~al}~\cite{#1}.}
	}

\newlength{\subLen}
\AtBeginDocument{\settoheight{\subLen}{$x$}}
\newcommand{\toSubLen}[1]{\resizebox{!}{\subLen}{#1}}
\newcommand\0{\toSubLen{0}}    
\newcommand{\X}{\textsc{x}}
\newcommand{\Y}{\textsc{y}}
\newcommand{\Z}{\textsc{z}}

\newcommand{\hatd}[1]{\hat{#1}^{\dag}}
\newcommand{\creop}{\hatd{a}}
\newcommand{\annop}{\hat{a}^{\phantom{\dag}}}

\newcommand{\ketvac}{\ket{\mathrm{vac}}}
\newcommand{\gcorr}[2]{g^{\left(#1\right)}\left(#2\right)}

\begin{document}

\title{Polarisation oscillations in birefringent emitter-cavity systems}

\author{Thomas D. Barrett}
\author{Oliver Barter}
\author{Dustin Stuart}
\author{Ben Yuen}
\author{Axel Kuhn}
\email{axel.kuhn@physics.ox.ac.uk}
\affiliation{University of Oxford, Clarendon Laboratory, Parks Road, Oxford  OX1 3PU, UK}

\date{\today}


\begin{abstract}
We present the effects of resonator birefringence on the cavity-enhanced interfacing of quantum states of light and matter, including the first observation of single photons with a time-dependent polarisation state that evolves within their coherence time.  A theoretical model is introduced and experimentally verified by the modified polarisation of temporally-long single photons emitted from a \Rb{} atom coupled to a high-finesse optical cavity by a vacuum-stimulated Raman adiabatic passage (V-STIRAP) process.  Further theoretical investigation shows how a change in cavity birefringence can both impact the atom-cavity coupling and engender starkly different polarisation behaviour in the emitted photons.  With polarisation a key resource for encoding quantum states of light and modern micron-scale cavities particularly prone to birefringence, the consideration of these effects is vital to the faithful realisation of efficient and coherent emitter-photon interfaces for distributed quantum networking and communications.
\end{abstract}


\maketitle


Cavity quantum electrodynamics (CQED) allows for the nature of light and matter to be interrogated through the enhanced interaction of an emitter with the resonant modes of a cavity~\cite{haroche13, brune96, gleyzes07}.  This allows these fundamental interactions to be leveraged for quantum technologies~\cite{kimble08, monroe02, monroe14, reiserer15, fruchtman16} and, consequently, realising novel regimes in CQED has the potential to impact both foundational research and cutting-edge technological applications.  Single photons are fundamental particles, they possess no deeper substructure, therefore it is tempting to consider their properties to be similarly immutable.  However, CQED has shown photons to be a far richer resource, with a high degree of control demonstrated over the wavepackets~\cite{nisbet11}, frequency~\cite{legero04}, polarisation~\cite{wilk07} and phase~\cite{specht09} of temporally-long single photons.  Here, we report the first observation of a single-photon with a time-dependent polarisation state that evolves along its wavepacket.  Moreover, this effect arises from a system increasingly prevalent in the pursuit of scalable quantum technologies.

\begin{figure}[t]
\includegraphics{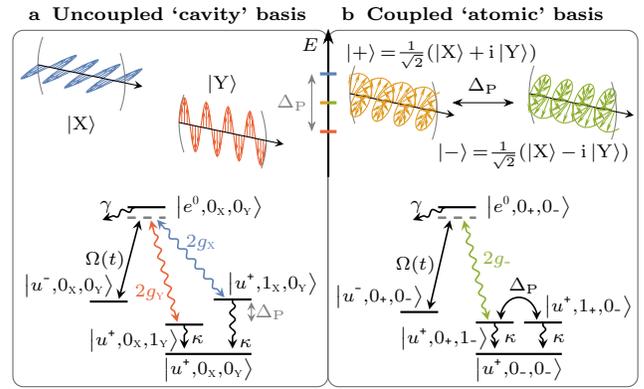}
\caption{Decomposition of cavity and $\Lambda$-system into coupled and uncoupled polarisation bases.  The upper plots show how the linear polarisation eigenmodes of a birefringent cavity can equivalently be considered as degenerate circularly polarised modes with an effective coupling between them.  The lower plots equivalently illustrate a simple $\Lambda$-system coupling of circularly polarised transitions within an atom in both bases.  The state notation is $\ket{x^{S}, n\sub{i}, n\sub{j}}$ with $x^{S}$ denoting an atomic state $x$ of spin $S$, and $n\subsc{z}$ the photon number in the cavity supporting mode $\ket{\mathrm{Z}}$.}
\label{fig:birefringenceModelSummary}
\end{figure}

\begin{figure*}[!t]
\includegraphics{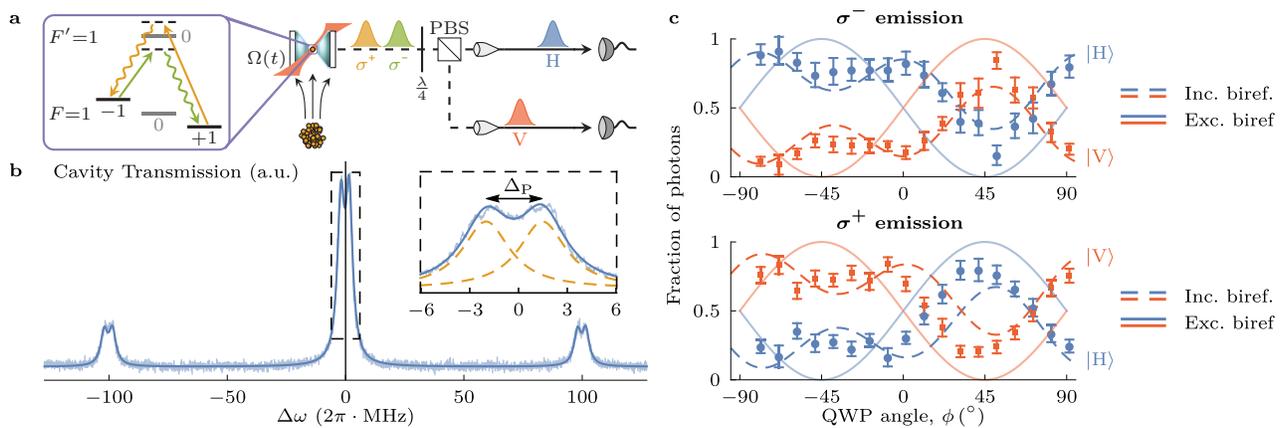}
\caption{Experimental summary.  (a)~Experimental setup for the production, routing and detection of polarised single photons.  The relevant couplings for each cavity-assisted Raman transition are distinguished by colour.  (b)~The transmission of laser light through the cavity for direct characterisation of cavity birefringence.  The cavity length is scanned over resonance with an incident laser which has sidebands at $\pm \SI{100}{\MHz}$ as a frequency reference.  The double-peaked Lorentzian (an adequate lineshape approximation for high-finesse cavities~\cite{sanchez16}) fit (solid blue) is comprised of the individual transmissions of the non-degenerate polarisation eigenmodes (dashed orange). (c)~Fractional routing of emitted photons as a function of the quarter-wave plate angle.  The dashed and solid theory traces include and exclude the effects of cavity birefringence respectively.  The error bars in both plots are found from the ${\pm}\sqrt{N}$ uncertainty on $N$ events exhibiting Poissonian counting statistics.}
\label{fig:experimentalSummary}
\end{figure*}

The coherent interfacing of light and matter qubits lies at the heart of many quantum networking proposals~\cite{kimble08,monroe02, monroe14, reiserer15, fruchtman16}, and the interaction of atom-like emitters with a single photonic mode of a resonator provides a platform for realising this control.  CQED is a vibrant field with single atoms and ions particularly suitable candidates with which to realise network nodes and single-photon sources due to their inherently homogeneous nature.  The \emph{a priori} deterministic emission of single photons into well-defined quantum states has been realised in both atom-cavity \cite{mckeever04, kuhn02, wilk07, vasilev10, nisbet13} and ion-cavity systems \cite{keller04}.  Proof-of-principle quantum networking demonstrations have leveraged this control to, for example, remotely entangle two atoms \cite{ritter12} and perform two-bit quantum gates \cite{reiserer14, hacker16, welte18}.  Improving the efficiency and scalability of such systems ultimately requires increasing the strength and reliability of the emitter-cavity coupling, motivating the development of microcavity resonators with tightly confined optical modes.  Micron-scale Fabry-Perot cavities, such as those formed between laser-ablated mirrors on the tips of optical fibres~\cite{steinmetz06, hunger10, gallego16}, provide open access to the mode for ease of coupling and the trapping of single atoms \cite{gallego18} or ions \cite{steiner13, keller17, ballance17}.  Moreover, work with Fabry-Perot microcavities has also demonstrated the enhanced coupling of light to molecules \cite{toninelli10} and to a variety of solid-state emitters including nitrogen-vacancy centers \cite{albrecht13, kaupp16}, quantum dots \cite{muller09, sanchez13, snijders18}, carbon nanotubes \cite{hummer16, jeantet16} and opto-mechanical devices \cite{flowers12, kashkanova16, zhong17}.  However, on these length scales the tightly-curved mirrors are highly susceptible to birefringence \cite{gehr10, hunger10, brandstatter13, takahashi14, mader15, garcia18} -- a lifting of the degeneracy of the two polarisation eigenmodes of the cavity -- due to the elliptical curvature of the mirrors \cite{uphoff15}.  Polarisation both strongly effects the interaction between light and atomic emitters, and is a potential basis for quantum information protocols \cite{wilk07b, reiserer14}, which has motivated the attempts to control this ellipticity-induced birefringece \cite{takahashi14, garcia18}. 	More generally the effects of birefringence on light incident on cavities has been studied in ringdown spectroscopy \cite{huang08, dupre15}, high-precision polarimeters \cite{ejlli18} and even for cavity-stabilisation proposals \cite{asenbaum11}.

In this Letter we present the first investigation of the interaction of quantum states of light and matter within a birefringent Fabry-Perot resonator.  We use a single \Rb{} atom strongly coupled to a birefringent cavity and observe the dynamic change in polarisation of the single photons emitted by a vacuum-stimulated Raman adiabatic passage (V-STIRAP) process \cite{kuhn02, kuhn09, wilk07c}.  Our experiment is uncommonly suited to this task as our cavity exhibits non-negligible polarisation-mode splitting despite being constructed with macroscopic mirror substrates -- a technology that allows for the reliable coupling of atoms to the cavity.  We begin, however, with a simple theoretical description of a birefringent atom-cavity system.



We decompose the cavity into a pair of orthogonal polarisation modes.  These can be the non-degenerate polarisation eigenmodes which independently couple to the atom, the so-called `cavity' basis, or the pair of polarisations which corresponds to the atomic transitions, the `atomic' basis.  \Cref{fig:birefringenceModelSummary}a summarises the system in the cavity basis for the extreme case where linearly polarised cavity eigenmodes couple circularly polarised atomic transitions.  A photon is emitted into a superposition of the cavity eigenmodes, and these eigenmodes accumulate a phase difference at a rate $\Delta\sub{P}$, the energy difference between them.  This results in a time-dependent oscillation between any pair of orthogonal polarisation states other than the cavity eigenmodes themselves.  Viewed in the atomic basis, \cref{fig:birefringenceModelSummary}b, the photon is emitted into only one of the considered polarisation states, with this oscillation then coupling the emitted state to its orthogonal counterpart.

Our approach can be formalised as an extension of the Jaynes-Cummings model and these details can be found in the Supplemental Material~\footnote{See Supplemental Material at \url{todo}, which includes Refs. \cite{kuhn10,kuhn15,johansson13,brasil13,pearle12,hong87,legero06}, for details of how the Jaynes-Cummings model is extended to consider a birefringent cavity and a characterisation of the single-photon source.}.
\nocite{kuhn10,kuhn15,johansson13,brasil13,pearle12,hong87,legero06}


At the heart of our experimental investigation is a cavity of non-negligible birefringence.  The cavity is \SI{339.289\pm0.002}{\um} long with a measured finesse of $\Fin{} = \num{117800\pm200}$.  The mirrors have a \SI{5}{\cm} radius of curvature and a ${\sim}\SI{1.5}{\mm}$ diameter.  Imbalanced mirror transmissions of ${\leq}\SI{1.6}{ppm}$ and ${\sim}\SI{40}{ppm}$ give a directional emission of the photons.  \Cref{fig:experimentalSummary}b shows a direct measurement of the cavity transmission from which we find two polarisation eigenmodes, each with a linewidth of $\Delta\omega\sub{FWHM}/2\pi=\SI{3.543\pm0.006}{\MHz}$, split by $\Delta\sub{P}/2\pi=\SI{3.471\pm0.004}{\MHz}$.  These eigenmodes are elliptically polarised with $\ketrm{X} = 0.888\ketrm{H} + 0.459 \e^{-2.709\ii}\ketrm{V}$ and $\ketrm{Y}$ correspondingly orthogonal.  In this work the cavity is tuned such that the desired resonance is between these two eigenmodes (\ie{} at $\Delta\omega=\SI{0}{\MHz}$ in \cref{fig:experimentalSummary}b).  The coupling parameters of the system are then $\{g,\kappa,\gamma\}/2\pi = \{4.77,1.77,3.03\}\,\si{\MHz}$, where $\kappa$ is the cavity field decay rate and $\gamma$ is the atomic amplitude decay rate, which places the experiment in the strong-coupling regime~\cite{kimble98}.

\def\foottext{This barred notation indicates that the angular dependence of each transition has not been included, which we consider explicitly due to the asymmetrically shifted coupling strengths in our system that are a result of the non-negligible magnetic field (as detailed in~\cite{barrett18b}).  For example, the Rabi frequency of the $\ket{F{=}1, m_{F}{=}{\pm}1} \leftrightarrow \ket{F\dash{=}1, m_{F}\dash{=}0}$ couplings in zero magnetic field is $\sqrt{5/24}\times\overline{\Omega}$, with the appropriate pre-factor for different transitions readily available in standard reference books~\cite{steck87}.  The quoted atom-cavity coupling rate, $g/2\pi = \SI{4.77}{\MHz}$, similarly assumes the zero-field transition strengths.  The laser-cavity Raman resonance was experimentally optimised to $2\pi\times\SI{7.5}{\MHz}$ above the zero-field $\ket{F{=}1,m_{F}{=}0}$ to $\ket{F\dash{=}1,m_{F}\dash{=}0}$ transition frequency as necessitated by the nonlinear Zeeman effects~\cite{barrett18b}.}

Each experimental cycle begins by loading \Rb{} atoms into a magneto-optical trap (MOT) ${\sim}\SI{8}{\mm}$ below the cavity for ${\sim}\SI{500}{\ms}$.  Atoms are then stochastically delivered into the cavity mode by an atomic fountain, which launches the MOT upwards at a velocity of ${\sim}\SI{1}{\meter \per \second}$.  The cloud is kept at a sufficiently low density such that we can consider only zero or one atom to be in the cavity at any one time.  Polarised single photons are produced using a V-STIRAP process, summarised in \cref{fig:experimentalSummary}a, between the $\ket{F{=}1, m_{F}{=}{\pm}1}$ ground state magnetic sublevels of the \DTwo{} line \cite{wilk07c, wilk07}.  An external magnetic field aligned along the cavity axis lifts the degeneracy of these sublevels by $2\pi{\times}\SI{26}{\MHz}$, allowing a pump laser and the cavity to form a $\Lambda$-system with the $\ket{F\dash{=}1,m_{F\dash}{=}0}$ excited state.  When the pump is detuned from the cavity resonance by ${\pm}2\pi{\times}\SI{26}{\MHz}$, a Raman-resonant transition from $\ket{m_{F}{=}{\pm}1}$ to $\ket{m_{F}{=}{\mp}1}$ emits a $\sigma^{\pm}$ photon into the cavity.  As the atoms traverse the cavity mode, \num{20000} alternately-detuned pump pulses -- each with a $T=\SI{333}{\ns}$ long $\sin{}^{4}(t/T)$ intensity profile and a peak Rabi frequency of $\overline{\Omega}/2\pi=\SI{10.0}{\MHz}$ \footnote{\foottext} -- attempt to produce a stream of alternately-polarised photons at a repetition rate of ${\sim}\SI{1.5}{\MHz}$.  A single atom takes ${\sim}\SI{60}{\us}$ to transit the mode, with waist $\omega_0{\sim}\SI{26.8}{\um}$, which corresponds to more than \num{100} photon production attempts.  An atom can be considered to be effectively stationary -- and thus the atom-cavity coupling unchanged -- within the duration of a single pump pulse.  The pump laser is linearly polarised and injected orthogonally to the cavity mode such that it decomposes into an equal superposition of $\sigma^{+}$ and $\sigma^{-}$ light in the cavity basis.  Single photons are detected by superconducting nanowire detectors \footnote{Photon Spot, model number NW1FC780}. The dark count rates range from 5 to 66 per hour across the battery of detectors, and are thus negligible.  Every detection event is recorded at run-time with \SI{81}{\ps} precision by a time-to-digital converter \footnote{Qutools quTAU}.  A characterisation of the produced photons -- detailing their singular nature and coherence -- can be found in the Supplemental Materials~\cite{Note1}.

\def\footnotefit{The reduced chi-squared parameters \cite{hughes10} for the theoretical fits to the measured data for $\sigma^{\minus}$ emission in \cref{fig:experimentalSummary}c are $\chi^{2}_{\nu} = 1.1$ ($\chi^{2}_{\nu} = 11.0$) for $\ketrm{H}$ and $\chi^{2}_{\nu} = 2.3$ ($\chi^{2}_{\nu} = 23.7$) for $\ketrm{V}$ for the fits including (excluding) cavity birefringence effects.  An equivalent analysis for $\sigma^{\plus}$ emission from the atom gives $\chi^{2}_{\nu} = 5.7$ ($\chi^{2}_{\nu} = 14.4$) for $\ketrm{H}$ and $\chi^{2}_{\nu} = 8.4$ ($\chi^{2}_{\nu} = 18.0$) for $\ketrm{V}$.  Systematic effects due to optical elements beyond the cavity were verified to be negligible with the polarisation of light passing through both vacuum chamber viewports rotated by ${<}\SI{1.3}{\percent}$.  The purity of the photon polarisation states exceeds the best observed routing of $\SI{91}{\percent}$ (achieved at $\phi=\SI{-68.5}{\degree}$ for $\sigma^{-}$ emission), as this is averaged over the entire length of the time-dependent photon states and achieved with only a quarter-wave plate.}

We observe polarisation states of photons emitted from the cavity that are significantly modified from those initially emitted by the atom.  A polarisation analyser consisting of a quarter-wave plate and a polarising beamsplitter (PBS) split the photon stream into two paths prior to detection.  \Cref{fig:experimentalSummary}c shows the fractional routing of the photons as a function of the quarter-wave plate angle.  The measured behaviour coincides well with the model including birefringence effects (dashed traces) and is in disagreement with the simple prediction of the emission of circularly polarised photons (solid traces) \footnote{\footnotefit}, as would be expected in the case of negligible cavity birefringence.

\begin{figure}[t]
\includegraphics{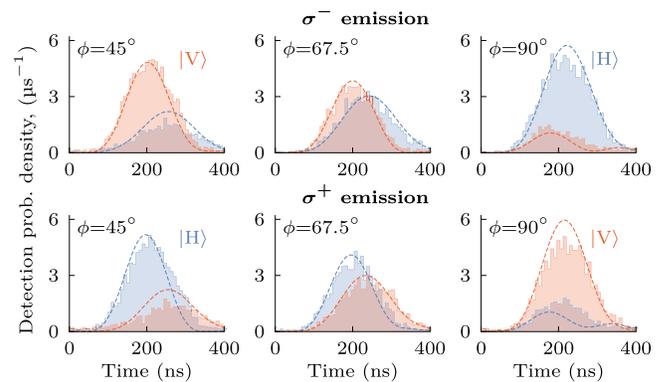}
\caption{Time-dependent polarisation of the emitted photons.  Measured data (solid traces, filled) overlaid with theoretical predictions (dashed traces) of the photon wavepackets measured in various polarisation bases using a quarter-wave plate at angle $\phi$ and a PBS.  The blue and red wavepackets correspond to detections in the different outputs of the PBS and so correspond to different polarisations.  The measured data is presented as density histograms with bin widths of \SI{8}{\ns}.}
\label{fig:photonWavepacketsVsQWP}
\end{figure}

The extreme length of the photons (\SI{333}{\ns}, as determined by the duration of a pump pulse, in comparison to the ${<}\SI{100}{ps}$ timing jitter of the detectors) allows us to examine how the polarisation changes with time and so observe polarisation oscillations within a single photon's wavepacket.  \Cref{fig:photonWavepacketsVsQWP} shows time-resolved distribution of photon detections at each detector for three orientations of the routing quarter-wave plate.  This corresponds to a measurement of the polarisation state along the photon wavepacket.  The differing wavepacket profiles measured at each detector show that the polarisation state is changing along the photon length.  This is a result of the birefringence-induced coupling between the two orthogonal polarisation states onto which the photon is projected by our measurement.  Only if we were measuring in the (uncoupled) cavity basis would the relative population of each polarisation mode be unchanged along the photon length.  This can again be contrasted to the expected behaviour in the case of negligible cavity birefringence where static photon polarisations would be emitted from the cavity. The photon counts at each detector would then be some constant fraction of the overall wavepacket, regardless of the chosen measurement basis.

\begin{figure}[t]
\includegraphics{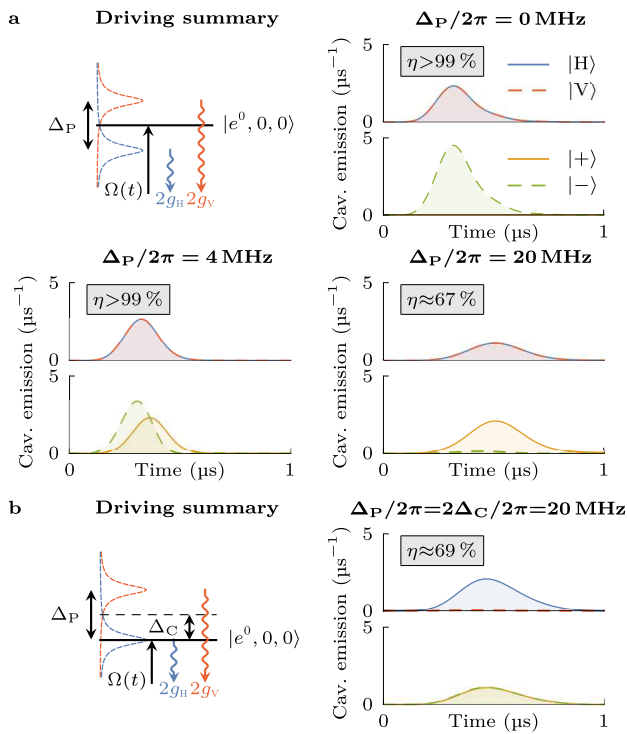}
\caption{Polarisation dynamics of emitted photons simulated for increasing splitting of the linearly polarised cavity eigenmodes, $\ketrm{H}$ and $\ketrm{V}$.  Photon emission is driven by a \SI{1}{\us} pump pulse with a ${\sin}^{4}$ intensity profile and a peak Rabi frequency $\Omega/2\pi{=}\SI{7}{\MHz}$ and $\Omega/2\pi{=}\SI{2}{\MHz}$ for $\Delta\sub{P}/2\pi{=}\{0,4\}\si{\MHz}$ and $\Delta\sub{P}/2\pi{=}\SI{20}{\MHz}$ respectively.  The cavity resonances are tuned such that, (a) the polarisation eigenmodes are oppositely detuned from Raman resonance, or, (b) a single polarisation eigenmode is on Raman resonance.  The photon wavepacket is plotted in both a linear and circular polarisation basis, with the overall efficiency of the emission process, $\eta$, inset.
}
\label{fig:birefringenceLimitingCases}
\end{figure}

Having experimentally observed that cavity birefringence modifies the polarisation of emitted photons, it is natural to consider some general limiting cases to further examine these effects.  To isolate only the effects of birefringence we return to the three-level system shown in \cref{fig:birefringenceModelSummary} and consider the emission of $\sigma^{\minus}$ photons.  Realistic coupling parameters are chosen, $\{g,\kappa,\gamma\}/2\pi=\{4,2,0\}\si{\MHz}$, disregarding spontaneous emission.  Whilst any physical system will be subject to this decay, in general it is a loss leading to an incoherent evolution which we ignore.  If the cavity eigenmodes are aligned with the circularly polarised atomic basis, the system reduces to a simple three-level model with a single cavity coupling and birefringence has no effect.  Therefore we consider the case of minimal overlap between the cavity and atomic bases -- a `linear' cavity with $\{\ketrm{X},\ketrm{Y}\}=\{\ketrm{H},\ketrm{V}\}$.

\Cref{fig:birefringenceLimitingCases}a summarises the emissions with the birefringent cavity modes oppositely detuned from Raman resonance with the pump laser.  With no birefringence ($\Delta\sub{P}/2\pi {=} \SI{0}{\MHz}$) the photon is emitted in the expected $\ketrm{-}$ mode with close to a unity efficiency.  As has already been seen in the measured output of our physical system, a birefringence that is comparable to the other coupling rates of the system (we take $\Delta\sub{P}/2\pi{=}\SI{4}{\MHz}$ for our simulated system) results in a time-dependent polarisation state of the emitted photon due to the coupling between the $\ketrm{+}$ and $\ketrm{-}$ modes.  Increasing the birefringence further (to $\Delta\sub{P}/2\pi{=}\SI{20}{\MHz}$) we find that the cavity emission has flipped and is almost entirely in the $\ketrm{+}$ mode, which is orthogonal to the polarisation originally emitted by the atom.  This striking effect can be understood as the adiabatic elimination of $\ket{u^{\plus},0_{\plus},1_{\minus}}$ in the couplings $\ket{u^{\minus},0_{\plus},0_{\minus}} \leftrightarrow \ket{u^{\plus},0_{\plus},1_{\minus}} \leftrightarrow \ket{u^{\plus},1_{\plus},0_{\minus}}$ (where we have implicitly considered the adiabatic elimination of $\ket{e^{\0},0_{\plus},0_{\minus}}$ in the photon production process).

The emission efficiency is also reduced by increased birefringence due to the weakened coupling of the atom to the off-Raman-resonant cavity modes.  For our system this necessitates a correspondingly weaker pump pulse, otherwise the dominant effect becomes Rabi oscillations between $\ket{u^{\minus}}$ and $\ket{e^{\0}}$.  The efficiency is still reduced if one of the cavity eigenmodes is set to be Raman-resonant, however for sufficiently large birefringence the cavity emission is then almost entirely into this linear polarisation mode (see \cref{fig:birefringenceLimitingCases}b).  Any cavity, regardless of the splitting or polarisation of its eigenmodes, can in principle couple any transition that is not mutually orthogonal to both eigenmodes, as there will always be at least one cavity mode that decomposes to have some contribution of the desired polarisation component.   In this case the cavity can then be understood to act as a filter, transmitting only the polarisation of light it is capable of supporting.  In the paraxial approximation, this precludes $\pi$-polarised modes because their field vector points along the cavity axis.


The modified photon polarisations and coupling strengths we have observed have the potential to impact a wide array of cavity-based schemes.  For example, in systems where information is encoded into the polarisation state of single photons these effects could lead to a loss of coherence and increased error rates.  Additionally, birefringence will result in a distinguishability between different emitter-cavity nodes, even when using inherently homogenous emitters such as atoms or ions.  Even supposing that preserving the polarisation state of the light is not required, there is still a reduction in coupling strength to any mode not aligned with the cavity eigenmodes.

We foresee the effects and model presented in this work as guiding the on-going efforts towards minimising and tailoring birefringence in high-cooperativity cavities.  This will be essential to future experiments using cavity-enhanced interactions such as the pursuit of scalable quantum network architectures.


\begin{acknowledgements}
The authors would like to acknowledge support for this work through the quantum technologies programme (NQIT hub), and express gratitude for their helpful discussions with A.~Beige and to E.~Kassa and T.~Doherty for their proof-reading efforts.
\end{acknowledgements}

\bibliographystyle{apsrev4-1}
\bibliography{Birefringence_References}

\cleardoublepage

\begin{section}{Supplemental Material}

\subsection{Details of the model}

\subsubsection*{Setting up the Hamiltonian}

To produce single photons \emph{a priori} deterministically from a coupled atom-cavity system we use a V-STIRAP process~\cite{kuhn02,kuhn10}.  A pump laser together with the cavity form a Raman resonance between two ground levels in a three-level $\Lambda$-system.  In this there exists a `dark' eigenstate where the atomic population is distributed only across these two ground states, with the population of each given by the relative strengths of the couplings to the excited state.  The extended Jaynes--Cummings model describes such a system~\cite{kuhn09, kuhn15} and here we further extend this approach to consider the coupling of an atom to a birefringent cavity.  Energy level diagrams of the `three-level' $\Lambda$-system, considering photon number states in either the cavity or atomic bases, are shown in Fig.\ 1 of the main text.

Consider a birefringent cavity with its two non-degenerate eigenmodes, $\{\ketrm{X}, \ketrm{Y}\}$.  Introducing an atom coupled simultaneously to these cavities, our state notation is $\ket{s, n\sub{\X}, n\sub{\Y}}$ where $s$ is the atomic state, and $n\sub{\Z}$ is the photon number in the cavity supporting mode $\ketrm{Z}$.  The dimensions of our state space is $M \times N \times N$ where we consider $M$ atomic states and restrict each cavity to have $0,1,\dots,N-1$ photons within it at any time.  This work considers single-photon generation and so we choose $N=2$.

The Hamiltonian for the coupled atom-cavity system can be written as
\begin{equation}
\Ham = \Ham\sub{atom} + \Ham\sub{cav} + \Ham\sub{int}.
\label{eq:Ham}
\end{equation}
The first two terms, corresponding to the energy of the bare atom and the bare cavity modes, are given by
\begin{align}
\Ham\sub{atom} &=\hbar \sum\limits_{s} \omega\sub{s} \ket{s}\bra{s}, \label{eq:Hatom} \\
\Ham\sub{cav} &= \hbar (\omega\sub{\X} \hatd{a}\sub{\X}\hat{a}\sub{\X} + \omega\sub{\Y} \hatd{a}\sub{\Y}\hat{a}\sub{\Y} ), \label{eq:Hcav}
\end{align}
where $\omega\sub{\X} - \omega\sub{\Y} = \Delta\sub{P}$ and $\hat{a}\sub{\X}$ and $\hat{a}\sub{\Y}$ are the annihilation operators for each cavity.  The laser and cavity interactions take the form
{\allowdisplaybreaks 
\begin{align}
\Ham\sub{int} &= -\hbar ( \Ham_\mathrm{int,L} + \Ham_{\mathrm{int,C}} ), \label{eq:Hint}\\
\Ham_\mathrm{int,L}&= \frac{\Omega(t)}{2} 
	\left( \sum\limits_{ s_{u},s_{e} } \ket{s_{e}}\bra{s_{u}} \e^{-\ii \omega\sub{L}t} + \ket{s_{u}}\bra{s_{e}} \e^{\ii \omega\sub{L}t} \right), \label{eq:HintL}\\
\Ham_{\mathrm{int,C}}&= g
	\left( \sum\limits_{ s_{u},s_{e} } \ket{s_{e}}\bra{s_{u}} \hat{a}_{k} + \hatd{a}_{k} \ket{s_{u}}\bra{s_{e}} \right), \label{eq:HintC}
\end{align}
}where $\omega\sub{L}$ is the laser frequency, $g$ is the atom-cavity coupling rate, $s_{u}$ and $s_{e}$ are respectively the ground and excited atomic states coupled by each.  The annihilation operator $\hat{a}_{k}$ is for the appropriate polarisation that couples the atomic transition but, practically, it must be expressed in the cavity basis (as this is the basis of the model, chosen such that model cavities are uncoupled).

\subsubsection*{Switching the polarisation bases of the model}

It is straightforward to express the appropriate photon ladder operators that couple an atomic transition in the atomic polarisation basis -- which we take to be circularly polarised, $\{\ket{+},\ketrm{-}\}$ -- however this basis is typically different from that of the birefringent cavity modes, $\{\ketrm{X},\ketrm{Y}\}$.  A known reference with respect to which these can be found is the linearly polarised lab basis, $\{\ketrm{H},\ketrm{V}\}$.  Describing polarised light using Jones calculus we move between different bases using rotation matrices of the form
\begin{equation}
	\hat{R}_{ij} = \mqty(\e^{\ii \phi_{1,ij}} \, \alpha_{ij} & -\e^{-\ii \phi_{2,ij}} \, \sqrt{1-\alpha_{ij}^{2}} \\  \e^{\ii \phi_{2,ij}} \, \sqrt{1-\alpha_{ij}^{2}} & \e^{-\ii \phi_{1,ij}} \, \alpha_{ij} ).
	\label{eq:genRotMat}
\end{equation}
where $\hat{R}_{ij}$ maps from basis $i$ to $j$, $0 \leq \alpha_{ij} \leq 1$ and $\alpha_{ij},\phi_{1,ij}, \phi_{2,ij} \in \mathds{R}$.  These mappings are unitary and thus $\hat{R}_{ji}=\hat{R}^{\dag}_{ij}$.

To correctly populate the model's states -- which, we recall, are in the cavity basis -- with photons produced from the atomic V-STIRAP transition, the creation operators for a photon in the atomic modes, $\creop\sub{\plus}$ and $\creop\sub{\minus}$, must be expressed in terms of the cavity creation operators, $\creop\sub{\X}$ and $\creop\sub{\Y}$.  The operator in the cavity basis that produces a photon in, for example, the $\ketrm{+}$ mode of the atomic basis can be written as
\begin{equation}
	\creop\sub{\plus} \ketvac\sub{C} = \ketrm{+}\sub{C} = \hat{R}\sub{AC}\ketrm{+}\sub{A}
\end{equation}
where the subscript on each state denotes the basis in which it would be written -- for example $\ketrm{+}\sub{A} \equiv \mqty(1 \\ 0)\sub{A}$.  Using \cref{eq:genRotMat} this becomes
\begin{align}
	& 
	\begin{aligned}
		\creop\sub{\plus} \ketvac\sub{C} = & \e^{\ii \phi\sub{1,AC}} \, \alpha\sub{AC} \, \ketrm{X}\sub{C} + \\
	  	& \e^{\ii \phi\sub{2,AC}}  \, \sqrt{1-\alpha\sub{AC}^{2}} \, \ketrm{Y}\sub{C}
	\end{aligned} \\
	\implies{} &\creop\sub{\plus} = \e^{\ii \phi\sub{1,AC}} \, \alpha\sub{AC} \, \creop\sub{\X}+  \e^{\ii \phi\sub{2,AC}}  \, \sqrt{1-\alpha\sub{AC}^{2}} \, \creop\sub{\Y}. \label{eq:creationOpR}
\end{align}

It only remains to find the rotation from the atomic to the cavity basis, which is equivalent to finding the rotation from each to the lab basis.  The atomic basis can be expressed in terms of the linear lab basis in the normal way,
\begin{equation}
	\ketrm{+}\sub{L} = \frac{1}{\sqrt{2}} ( \ketrm{H}\sub{L} + \ii \ketrm{V}\sub{L} ), \quad
	\ketrm{-}\sub{L} = \frac{1}{\sqrt{2}} ( \ketrm{H}\sub{L} - \ii \ketrm{V}\sub{L} ),
\end{equation}
which gives 
\begin{equation}
	\hat{R}\sub{AL} = \hatd{R}\sub{LA} = 
		\frac{1}{\sqrt{2}}\mqty(1 & \ii \\
								\ii & 1 ).
\label{eq:basisDecompositionInLab}
\end{equation}

As an example, the a simple theoretical system that was discussed in the main text is a cavity with linearly polarised eigenmodes ($\{\ketrm{X}{=}\ketrm{H},\ketrm{Y}{=}\ketrm{V}\}$).  In this case the appropriate mapping of the ladder operators between the atomic and cavity basis is then
\begin{align}
	\creop_{\plus} &= \tfrac{1}{\sqrt{2}}(\creop\subsc{x} + \ii\,\creop\subsc{y}), \\
	\creop_{\minus} &= \tfrac{1}{\sqrt{2}}(\creop\subsc{x} - \ii\,\creop\subsc{y}).
\end{align}
When considering the emission of $\sigma^{\minus}$ photons from this system, as is the case for figures 1 and 4 of the main text, equation (S6) becomes
\begin{equation}
	\Ham_{\mathrm{int,C}}= g\big( \ket*{e^{\0}}\bra*{u^{\plus}} \hat{a}\sub{\minus} + \hatd{a}\sub{\minus} \ket*{u^{\plus}}\bra*{e^{\0}} \big),
\end{equation}
where we have followed the state labelling convention from these figures by setting $\ket{s_{u}}=\ket{u^{\plus}}$ and $\ket{s_{e}}=\ket{e^{\0}}$.

To model the performance of a physical experiment, the orientation of the birefringent cavity eigenmodes can be measured in the lab by directly observing the transmission of polarised light through the cavity -- specifically by finding the polarisation uniquely coupled into just one of the modes.  For the cavity used in this work it was found that $( \alpha\sub{CL}, \phi\sub{1,CL}, \phi\sub{2,CL} ) =( 0.888, \SI{115.1}{\degree}, \SI{-40.1}{\degree} )$ -- which corresponds to elliptically polarised eigenmodes.
 
\subsubsection*{Simulating the Hamiltonian}

In this work the model is simulated by solving the master equation numerically using the \pkgfont{Qutip.mesolve} Python \pkgfont{package.function} \cite{johansson13}. This is done after the transformation of $\Ham{}$ into the rotating frame where it takes the form,
\begin{equation}
	\Ham\dash = \hatd{U} \cdot \Ham{} \cdot{} \hat{U} - \Ham{}_0, \qquad \hat{U} = \e^{- \ii \Ham{}_0 t},
\end{equation}
where $\Ham{}_0 = \diag{\Ham{}} = \Ham\sub{atom} + \Ham\sub{cav}$ is the `bare' Hamiltonian.  In this rotating frame $\Ham\dash{}$ has only zeros on the diagonal, which is to say the state basis is degenerate and time-dependence is included only in the off-diagonal coupling terms.  As we simulate the model in a rotating frame the photon creation operators, such as that in \cref{eq:creationOpR}, must, like all operators, be equivalently rotated into this frame before it is applied.

The master equation that models the time-evolution of this system, which has density matrix $\hat{\rho}$, is then \cite{brasil13,pearle12},
\begin{equation}
	\frac{\diffd}{\diffd t} \hat{\rho} = -\frac{\ii}{\hbar}\comm{\Ham\dash{}}{\hat{\rho}} + \hat{L}(\hat{\rho}),
	\label{eq:masterEquation}
\end{equation}
where $\hat{L}(\hat{\rho})$ is the Lindblad operator accounting for the relaxation of the system.  It takes the form
\begin{equation}
	\hat{L}(\hat{\rho}) = \sum\limits_{n} \left( 2\hat{C}_n \hat{\rho} \hatd{C}_n - \hat{\rho}\hatd{C}_n\hat{C}_n - \hatd{C}_n\hat{C}_n\hat{\rho} \right)
	\label{eq:lindbladOperator}
\end{equation}
with $\hat{C}_n$ the collapse operators.  In our case these couplings are either photon emission from the cavity, with $\sqrt{2\kappa} \annop_{\mathrm{\X,\Y}}$, or spontaneous decay between atomic levels $s_{e} \rightarrow s_{u}$ with $\sqrt{2\gamma_{eu}} * \ket{s_{g}}\bra{s_{e}}$, where $\kappa$ and $\gamma_{eu}$ are the decay rate of the electric field from the cavity and the decay rate of the atomic amplitude between levels $s_{e}$ and $s_{u}$, respectively.

The simulations presented in this work consider every coupling between all magnetic sublevels of the ground $\ket{F{=}1}$ and excited $\ket{F\dash{=}0,1}$ states.  Couplings to the other states of the excited manifold ($\ket{F\dash{=}2,3}$) are too far off-resonance to be significant.  Spontaneous decay to the $\ket{F{=}2}$ ground state is accounted for with the inclusion of an additional `dark' ground state in our model which atomic population can decay to but not return from.

\subsection{Characterising the single-photon source}

\begin{figure*}[t]
\includegraphics{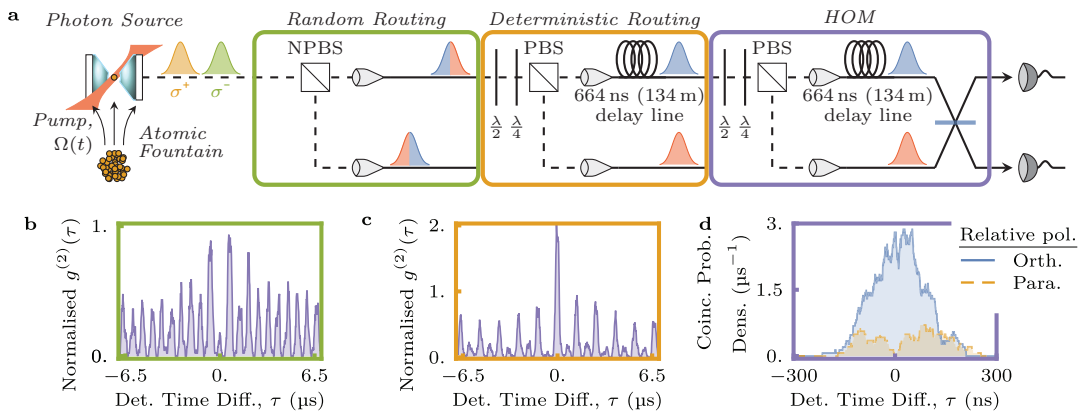}
\caption{Characterisation of the single-photon source. (a) Experimental setup showing the emission, routing and detection of a stream of single photons.  The three routing configurations correspond to three different experimental configurations used in this work.  (b)-(c) The second-order correlation functions, $\gcorr{2}{\tau}$, between cross-detector detection events for the `Random' and `Deterministic' routing configurations, respectively.  Both plots have bin widths of \SI{100}{\ns} and a pitch of \SI{20}{ns}.  (d)  Hong-Ou-Mandel interference of orthogonally (distinguishable) and parallel (indistinguishable) polarised photon pairs, shown as a sliding histogram with bin widths and pitch of \SI{40}{\ns} and \SI{4}{\ns} respectively.}
\label{fig:sourceCharacterisation}
\end{figure*}

The work presented in the main manuscript requires the photon-production process to provide single photons with coherent wavepackets spanning a sufficiently long time frame that the effects of birefringence upon its evolution can be observed.  Here we provide the experimental measurements to confirm and characterise these aspects of the source.  These measurements were carried out using the experimental set-ups detailed in \cref{fig:sourceCharacterisation}a.

The second-order correlation function, $\gcorr{2}{\tau}$, between between cross-detector events when randomly routing the emitted photons using a non-polarising beam splitter (NPBS) is shown in \cref{fig:sourceCharacterisation}b.  The correlation rate peaks at times corresponding to the \SI{664}{\ns} duty cycle of single-photon production, with each pump pulse \SI{300}{\ns} long (note that the experiments presented in the main manuscript used \SI{333}{\ns} long pump pulses). A suppressed central peak, corresponding to coincident detections within the same driving interval, is observed with $\gcorr{2}{0}=0.067$.  This non-zero probability is attributable to off-Raman-resonant processes resulting in multiple photon emissions within a single driving interval, however the low probability of these events in comparison to the emission of two photons from sequential driving intervals illustrates the single-photon nature of the source.  The non-zero possibility that a spontaneous emission during photon production leaves the atom in a `dark' state~\cite{barrett18b}, from which it can produce no more photons, results in reduced correlation rates at longer detection time differences.

To illustrate that the photon emissions are polarised, this experiment is repeated with a polarising beam splitter (PBS) and waveplates set to maximally route sequential emissions to different detectors (`Deterministic Routing').  An additional \SI{134}{\m} of optical fibre on one optical path delays one of these photons by the duty cycle of photon production such that sequentially emitted photons are delivered to the detectors simultaneously.  \Cref{fig:sourceCharacterisation}c shows that this routing is efficient -- and thus the emitted stream of photons is polarised -- with the large peaks separated by two driving intervals corresponding to the `correct' routing of sequential emissions.  The smaller peaks result from one photon being routed down the `wrong' path, which is unavoidable as the time-dependent polarisation state of the emitted photons (see Fig.\ 3 in the main text) means that no static polarisation optics can route sequential emissions down opposite paths with perfect efficiency.

The quantum interference of these photon pairs is characterised by a Hong-Ou-Mandel experiment \cite{hong87,legero04}.  This is achieved by interfering orthogonally (distinguishable) and parallel (indistinguishable) polarised photons on a 50:50 beam splitter.  The cross-detector coincidences as a function of detection time difference in these two cases are compared in \cref{fig:sourceCharacterisation}d.  For parallel polarised photons the suppression of these coincidences illustrates the `bunching' of indistinguishable pairs as they coalesce and exit into the same output mode.  The two-photon visibility is defined as the reduction in likelihood of measuring cross-detector coincidences for parallel polarised photons compared to the non-interfering orthogonally polarised reference.  Measured over the entire interaction time of the \SI{300}{\ns} long photons the visibility is \SI{70.8\pm4.6}{\percent}, which increases to $\geq\SI{97.8}{\percent}$ when considering only detections within less than $\SI{23}{ns}$ of each other.  This temporal variation in the photon distinguishability is a result of their coherence properties, the theory of which is described in detail in by Legero \emph{et al}~\cite{legero06}.  The behaviour observed in our system indicates the interference of narrowband photons with a $2\pi{\times}\SI{2.15}{\MHz}$ bandwidth.  A detailed consideration of the coherence properties of our photons can be found in our previous work~\cite{barrett18b}.

\end{section}

\end{document}